\documentstyle[12pt]{article}

\def\hybrid{\topmargin -20pt	\oddsidemargin 0pt
	\headheight 0pt	\headsep 0pt
        \textwidth 6.35in
        \textheight 9.65in
	\marginparwidth .875in
	\parskip 5pt plus 1pt	\jot = 1.5ex}

 \catcode`\@=11

\@addtoreset{equation}{subsection}
\@addtoreset{footnote}{section}
\def\theequation{\thesubsection.\arabic{equation}}

\newtoks\@stequation

\def\subequations{\refstepcounter{equation}%
  \edef\@savedequation{\the\c@equation}%
  \@stequation=\expandafter{\theequation}
  \edef\@savedtheequation{\the\@stequation}
  \edef\oldtheequation{\theequation}%
  \setcounter{equation}{0}%
  \def\theequation{\oldtheequation\alph{equation}}}

\def\endsubequations{\setcounter{equation}{\@savedequation}%
  \@stequation=\expandafter{\@savedtheequation}%
  \edef\theequation{\the\@stequation}\global\@ignoretrue
  \vspace*{-12pt} \\}

\hybrid
\def\baselinestretch{1.2}
\catcode`\@=11
\newcount\hour
\newcount\minute
\newtoks\amorpm
\hour=\time\divide\hour by60
\minute=\time{\multiply\hour by60 \global\advance\minute by-\hour}
\edef\standardtime{{\ifnum\hour<12 \global\amorpm={am}%
	\else\global\amorpm={pm}\advance\hour by-12 \fi
	\ifnum\hour=0 \hour=12 \fi
	\number\hour:\ifnum\minute<10 0\fi\number\minute\the\amorpm}}
\edef\militarytime{\number\hour:\ifnum\minute<10 0\fi\number\minute}

\def\draftlabel#1{{\@bsphack\if@filesw {\let\thepage\relax
   \xdef\@gtempa{\write\@auxout{\string
      \newlabel{#1}{{\@currentlabel}{\thepage}}}}}\@gtempa
   \if@nobreak \ifvmode\nobreak\fi\fi\fi\@esphack}
	\gdef\@eqnlabel{#1}}
\def\@eqnlabel{}
\def\@vacuum{}
\def\draftmarginnote#1{\marginpar{\raggedright\scriptsize\tt#1}}

\def\draft{\oddsidemargin -.5truein
	\def\@oddfoot{\sl preliminary draft \hfil
	\rm\thepage\hfil\sl\today\quad\militarytime}
	\let\@evenfoot\@oddfoot	\overfullrule 3pt
	\let\label=\draftlabel
	\let\marginnote=\draftmarginnote
   \def\@eqnnum{(\theequation)\rlap{\kern\marginparsep\tt\@eqnlabel}%
\global\let\@eqnlabel\@vacuum}  }

\catcode`\@=11
\def\section{\@startsection {section}{1}{0pt}{-3.5ex plus -1ex minus
 -.2ex}{2.3ex plus .2ex}{\raggedright\large\bf}}
\catcode`\@=12
\newskip\humongous \humongous=0pt plus 1000pt minus 1000pt

\newif\ifdtup

\def\s{\sigma}

\def\be{\begin{equation}}
\def\ee{\end{equation}}
\def\ba{\begin{eqnarray}}
\def\ea{\end{eqnarray}}
\def\bs{\begin{subequations}}
\def\es{\end{subequations}}

\def\s{\sigma}

\def\gappeq{\mathrel{\rlap {\raise.5ex\hbox{$>$}}
{\lower.5ex\hbox{$\sim$}}}}
\def\lappeq{\mathrel{\rlap{\raise.5ex\hbox{$<$}}
{\lower.5ex\hbox{$\sim$}}}}
\def\epeq{\mathrel{\rlap {\raise.5ex\hbox{$\sim$}}
{\lower.5ex\hbox{$-$}}}}

\def\str{{\rm string}}

\begin{document}
\renewcommand{\theequation}{\thesection.\arabic{equation}}
\newcommand{\beq}{\begin{equation}}
\newcommand{\eeq}[1]{\label{#1}\end{equation}}
\newcommand{\ber}{\begin{eqnarray}}
\newcommand{\eer}[1]{\label{#1}\end{eqnarray}}
\begin{titlepage}
\begin{center}

\hfill CERN-TH/95-224\\
\hfill LPTENS-95/42\\
\hfill hep-th/9509017\\

\vskip .2in

{\large \bf String Gravity and Cosmology: Some new ideas
\footnote{To appear in the proceedings of the Four Seas Conference,
Trieste, June 1995.}}
\vskip .4in

{\bf Elias Kiritsis and Costas Kounnas\footnote{On leave from Ecole
Normale Sup\'erieure, 24 rue Lhomond, F-75231, Paris, Cedex 05,
FRANCE.}}\\
\vskip
 .3in

{\em Theory Division, CERN,\\ CH-1211,
Geneva 23, SWITZERLAND} \\

\vskip .3in

\end{center}

\vskip 1.2in

\begin{center} {\bf ABSTRACT } \end{center}
\begin{quotation}\noindent
String theory provides the only consistent framework so far that
unifies
all interactions including gravity. We discuss gravity and cosmology
in string theory. Conventional notions from general relativity like
geometry, topology etc. are well defined only as low energy
approximations in string theory. At small distances physics deviates
from the field theoretic intuition.
We present several examples of purely stringy phenomena which imply
that the physics at strong curvatures can be quite different from
what one might expect
from field theory. They indicate new possibilities in the context of
quantum cosmology.
\end{quotation}
\vskip 3.0cm
CERN-TH/95-224\\
September 1995\\
\end{titlepage}
\vfill
\eject
\def\baselinestretch{1.2}
\baselineskip 16 pt
\noindent
\section{Introduction}
\setcounter{equation}{0}
\noindent
It is no accident that (super)string theory attracted the attention
of theorists during the last decade (for a review see \cite{gsw}).
It is the only theory we have so far, that contains a consistent
theory of quantum gravity, a feat impossible to reproduce using
conventional quantum field theory.

String theory is based on the idea that the elementary building
blocks
of matter, instead of being point-like particles (with local
interactions), are
strings (one dimensional objects, either closed or open\footnote{From
now on, for concreteness we will consider closed strings.}).
Conventional particles are identified with the eigenmodes of the
string.
Thus, the string vibrating in two different modes, corresponds to two
distinct particles.
Unlike field theory, there is automatically a mass scale inherent in
string
theory, namely the tension of the string. Since the theory always
turns out to contain a graviton, it is natural to identify this scale
with the Plank mass.

There are several attractive features of the theory. We list below
some of them.

$\bullet$  String theory is a finite theory at short distances. This
is due to the fact that string interactions are not localized at a
point.

$\bullet$  At large distances (much bigger than the Plank length
$\sim 10^{-33}$ cm) strings look like point-like objects and an
ordinary particle description is valid. Gauge symmetries and gravity
appear naturally at low energy.

$\bullet$  A consequence of the above is that string theory contains
a consistent (UV-finite) theory of gravity. Moreover, it
automatically unifies all interactions: gravitational, gauge and
Yukawa.
This, so far, is not possible to achieve in the context of field
theory.

$\bullet$  In order that string theory contains fermions, some form
of (broken) spacetime supersymmetry seems to be  needed. This is nice
since we know that (spontaneously broken) supersymmetry can help with
hierarchy type of problems.

The most useful formulation of the theory so far is the first
quantized formulation. Although there have been attempts to construct
string field theory,
the problem is far from solved. This explains our partial knowledge
of the full symmetry of the theory.

The first quantized formulation of field theory as developed by Dirac
and Feynman rests on representing field theory amplitudes as a sum
over paths of point particles
\be
\langle x|x'\rangle \sim \int_{0}^{T}d\tau\int_{x(0)=x}^{x(T)=x'}
[Dx(\tau)]~~e^{iS~[x(\tau)]}\label{field}
\ee
with
\be
S~[x(\tau)]=\int_{0}^{T}d\tau~\left[G_{\mu\nu}\dot x^{\mu}\dot
x^{\nu}
+A_{\mu}\dot x^{\mu}+\cdots\right]
\ee
where $\tau$ parametrizes the path, $G_{\mu\nu}$ is the metric,
$A_{\mu}$ a gauge field and the dots stand for other interactions.
The action here is that of a one-dimensional field theory defined
over the path of the particle.

In string theory, one can write a similar formula for the amplitude
for string propagation. A closed string, when it propagates
(classically) sweeps a
two-dimensional cylinder, (called the world-sheet) and the amplitude
now is a two-dimensional generalization of the field theory formula
(\ref{field}).
A closed string is a ring whose position is described by
$x^{\mu}(\sigma)$, where $0\leq \sigma\leq 1$ parametrizes the ring.
Then,
\be
\langle x(\sigma)|x'(\sigma)\rangle\sim
\int[Dx(\sigma,\tau)]~~e^{iS~[x(\sigma,\tau)]}\label{path}
\ee
\be
S~[x(\sigma,\tau)]={1\over 4\pi \alpha'}\int~d\sigma
d\tau~\left[G_{\mu\nu}(
\dot x^{\mu}\dot x^{\nu}-x'^{\mu}x'^{\nu})+B_{\mu\nu}(\dot
x^{\mu}x'^{\nu}-(\mu\to\nu))+\cdots\right]
\label{action}
\ee
where dot stands for derivative with respect to $\sigma$,
$G_{\mu\nu}$ is the metric, $B_{\mu\nu}$ is an antisymmetric tensor,
and the dots contain interactions with other massless fields (gauge
fields etc.)

Perturbative string theory contains two parameters.

$\bullet$ The first is the string tension $\alpha'$, which appears in
the action of the two-dimensional $\sigma$-model (\ref{action}).
It has dimensions of inverse mass-squared.
It sets the length (or mass) scale of the theory.
Note also that it is the coupling constant of the world-sheet
$\s$-model (\ref{path},\ref{action}) which describes string
propagation.
For small values of $\alpha'$ the $\s$-model is semiclassical.

$\bullet$ The second parameter is the string coupling constant
$g_{\str}$, which
is dimensionless, and governs the strength of string interactions.
It is the loop-expansion parameter of perturbative string theory.

Thus, $\alpha'$, the $\s$-model coupling constant, controls $stringy$
effects, in the sense that when $\alpha'
\to 0$ the theory becomes equivalent to a field theory.
On the other hand, $g_{\str}$, the string theory coupling constant,
controls
$quantum$ effects (when $g_{\str}\to 0$ the theory is classical).

At tree level, the Plank mass (or Newton's constant) is given in
terms of these two parameters as
\be
M^{2}_{\rm Planck}\sim {1\over g^{2}_{\str}~\alpha'}
\ee

Another interesting feature of (super)string theory is that, in
principle,
the dimension of spacetime can be any integer between zero and ten.
This gives the possibility that the theory determines dynamically the
dimension of spacetime to be four, although we do not understand the
mechanism so far.
All ground states with a four-dimensional spacetime contain some
universal fields: The metric (graviton) $G_{\mu\nu}$, the
antisymmetric tensor\footnote{In four dimensions $B_{\mu\nu}$ is
equivalent to a pseudoscalar,
usually called the axion.} $B_{\mu\nu}$ and a scalar field $\Phi$,
the dilaton.
It is interesting that the string coupling $g_{\str}$ is related to
the expectation value of the dilaton as
\be
g_{\str}=\langle e^{\Phi}\rangle
\ee
which indicates that the coupling constant of string theory, although
undetermined in perturbation theory, could be determined by
non-perturbative effects.

Since particles are in correspondence with the eigenmodes of the
string, it is obvious, that string theory describes the interactions
of an infinite number of
particles.
Some of them are massless. We mentioned already, that a flat
four-dimensional
ground state contains the ``universal excitations", $G_{\mu\nu},
B_{\mu\nu}, \Phi$ as well as gauge fields, fermions, and scalars
whose quantum numbers and
detailed interactions depend on the ground state.
The theory also contains towers of massive states, most of them
having masses
of the order of, or bigger than the Plank mass.

The classical equations of motion of string theory turn out be
equivalent with the conformal invariance of the two-dimensional
$\s$-model (\ref{path},\ref{action}).
Conformal invariance translates into the vanishing of the
$\beta$-functions
of the $\s$-model.
For small $\alpha'$, one can derive the $\s$-model one-loop
$\beta$-function equations and show that they come from the following
spacetime action\footnote{
To be distinguished from the world-sheet $\s$-model action
(\ref{action}).}
\be
S_{st}=M^{2}_{\rm Plank}\int \sqrt{G}e^{-2\Phi}\left[R+4(\nabla
\Phi)^2-
{1\over 12}H_{\mu\nu\rho}H^{\mu\nu\rho}+\cdots +{\cal
O}(\alpha')\right]
\label{st}\ee
with
\be
H_{\mu\nu\rho}=\partial_{\mu}B_{\nu\rho}+{\rm cyclic}~~{\rm
permutations}
\ee
We have displayed only the bosonic universal fields, $G_{\mu\nu}$,
$B_{\mu\nu}$, $\Phi$ (which we will call the gravitational sector).
The dots stand for the rest of the fields.
As indicated, there are higher order in $\alpha'$ (stringy)
corrections
with more than two derivatives. There are, for example,
$R_{\mu\nu\rho\s}
R^{\mu\nu\rho\s}$ terms.

This two-derivative effective action is valid when all relevant
length scales
are much smaller that the string scale $\sim 1/\alpha'$.
In particular, when curvatures become of the order of the string
scale, the $\alpha'$-corrections (neglected here) are large and we
cannot trust (\ref{st}).
The natural question to address in this context is: Can we handle
strong curvatures? Or, can we sum-up the $\alpha'$-expansion?

The answer to this question in given by the concept of Conformal
Field Theory.
It is a formalism which is exact (non-perturbatively) in $\alpha'$,
and thus
contains all (perturbative) stringy effects.
It should be thought of as an infinite dimensional analogue of fields
like the metric, the antisymmetric tensor, gauge and scalar fields
etc.
It also provides a handle on the perturbative  symmetries of string
theory.

Since string theory contains a massless graviton, we certainly
have diffeomorphism invariance. Massless gauge fields indicate the
presence of gauge invariance.

\noindent
Is the symmetry of string theory the direct product of such
field-theoretic symmetries?

\noindent
Although we do not know the symmetries of string theory in their full
glory, we certainly know that the answer to the previous question is
$negative$.
We can find extra symmetries that go beyond field theory.
Such symmetries come under the name of $duality$ symmetries.
We have examples where two $different$ effective field theory
solutions
correspond to the same string theory solution.
Duality symmetries act in a non-perturbative way on the
$\alpha'$-expansion.
In this respect, they are stringy symmetries.
What accounts for the difference here is that our probes (strings)
are
extended objects.

It turns out that concepts like geometry, topology, gauge symmetry,
dimension
of space etc. are low energy approximations in string theory.
In view of this we can ask: can every classical solution to string
theory be described completely in terms of the standard fields, like
the metric $G_{\mu\nu}$ etc.?
The answer turns out again to be $negative$.
We know of solutions which have semiclassical regions where geometry
is well defined, as well as regions that are ``fuzzy" (no geometric
description).
We also know of solutions where there is no semiclassical region at
all.

\noindent
When do we have a good geometrical description? This exists only if
solutions contain parameters that can be varied in such a way that
relevant scales (volume, curvature) become much larger than the
string scale.

\section{A Simple Example: String Theory on a Circle}
\setcounter{equation}{0}
\noindent
We will consider here a closed string moving on a manifold that
contains a
circle of radius $R$.
The part of the two-dimensional $\s$-model that describes the circle
has the following action:
\be
S_{\rm circle}={R^2\over 4\pi\alpha'}\int d^2
\xi~\partial_{\mu}\phi\partial^{\mu}\phi
\ee
where we have explicitly displayed the radius dependence of the
action and $\phi$ is a two dimensional field that takes values in
$[0,2\pi]$.
Note that $R$ has dimensions of length since the angle $\phi$ is
dimensionless.

In the field theory case, we know that for a particle moving on a
circle
the momentum is quantized
\be
p={m\over R}
\ee
where $m$ is an integer.
In this case if an internal dimension is such a circle the 4-D mass
of such
momentum excitations has a term proportional to $p^2$ (in the
Kaluza-Klein
framework.
Thus, in field theory
\be
M^2={m^2\over R^2}+\cdots\label{ft}
\ee
where the dots summarize other contributions.
Note that $p^2$ is the spectrum of the Laplacian operator ${1\over
R^2}{\partial^2\over \partial \phi^2}$ on the circle.
Knowing the spectrum of masses (or the Laplacian) we can reconstruct
the manifold.
Thus here geometry alone determines the spectrum and vice versa.
In particular by measuring the low lying spectrum we can measure the
size
(radius) of the circle.

In the string case, there are excitations that are not of the
momentum type.
The string can also wrap (several times) around the circle.
The ``energy" of such winding excitations depends on the string
tension $\alpha'$ and the total length of the winding string in a
very simple (linear)
fashion:
\be
E_{\rm winding}=~n~{R\over \alpha'}
\ee
where $n$ is the winding number (an integer) and $2\pi nR$ is the
total
length of the string.
Such excitations give additional contributions to the 4-D masses and
the full formula is
\be
M^2_{\rm string}={m^2\over R^2}+{n^2R^2\over \alpha'^2}+\cdots
\label{mass}
\ee
Imagine that we have a circle of large radius, $R^2>>\alpha'$. Then,
from (\ref{mass}), we observe that the low lying spectrum is given
solely in terms of momentum (field theory-like) excitations.
The states with non-zero winding numbers are comparatively very
massive,
namely of the order of the Plank mass.
Thus an observation of the low energy spectrum is described by
geometry
and we will measure the (large) value of the radius.

Consider however, the case where $R^2\sim \alpha'$.
Then the low lying spectrum contains both winding and momentum
excitations,
and in fact is not described by any (one-dimensional) geometry.
In this region, the geometric description breaks down.

Let us look now at the opposite limit, $R^2<<\alpha'$.
Now the low energy spectrum is composed solely in terms of the
winding excitations and low lying masses are
\be
M^2_{\rm low-lying}\epeq{n^2\over (\sqrt{\alpha'}/R)^2}\label{dmass}
\ee
Now this spectrum is similar to the field theory spectrum (\ref{ft})
, and there is again a geometric description in terms of a circle.
However, by comparing (\ref{ft}) and (\ref{dmass}) we will measure an
effective radius
\be
\tilde R={\sqrt{\alpha'}\over R}
\ee
which is obviously different from $R$.
We can thus conclude that
\be
R_{eff}\geq \sqrt{\alpha'}
\ee
and that there is an effective $minimum$ $size$ for the manifold in
string theory.
This minimum length, $\sqrt{\alpha'}$, is of the order of the Plank
length.

The above discussion follows from the observation that there is a
symmetry
in the stringy spectrum (\ref{mass}) of the circle, namely
\be
R\to {\sqrt{\alpha'}\over R}\label{tsd}
\ee
and a simultaneous exchange of winding and momentum excitations.
Such a symmetry is known as ``target space duality".
Another way to state the effect of this symmetry is: the two
$\s$-models associated to two $distinct$ geometries, namely a circle
of radius $R$ and a circle of radius $\sqrt{\alpha'}/R$ correspond to
the $same$ Conformal Field Theory, and thus to the same classical
string solution.
Target space duality symmetry is particular to the theory of strings
but not to field theory, since the necessary ingredient, namely
winding modes do not exist in field theory\footnote{Sometimes field
theories have effective excitations
which are stringy, for example Nielsen-Olesen vortices. In such cases
one could in principle have such a behavior.}.
It is obvious from (\ref{tsd}) that target space duality is $not$ a
symmetry order by order in $\alpha'$ (the $\s$-model coupling
constant).
We need the exact solution of the theory to see the symmetry.
Thus target space duality is a non-perturbative symmetry of the
$\s$-model.

Another relevant observation concerns the self-dual radius
$R=\sqrt{\alpha'}$. At this point, the symmetry of the $\sigma$-model
is enhanced from $U(1)_{L}\times U(1)_{R}$ to $SU(2)_{L}\times
SU(2)_{R}$.
Chiral $\s$-model symmetries, are associated to gauge symmetries in
string
theory.
Thus the theory with $R=\sqrt{\alpha'}$ has a $SU(2)\times SU(2)$
unbroken
gauge symmetry. When $R$ moves away from this value, the gauge
symmetry
is broken to $U(1)\times U(1)$.
In this respect, $R$ is like the expectation value of a Higgs field.

Although we indicated the simplest example of such a symmetry
\cite{ky}, it is more general, \cite{bu}-\cite{gr}.
It can be shown to be an exact symmetry order by order in string
perturbation theory \cite{k1}.
In general, there are two parts of the spectrum which we will
continue to call
momentum and winding modes, which are interchanged by target space
duality
The effective geometry seen by the momentum modes is in general
different
from the one seen by the winding modes (the dual geometry).

\section{Implications for Effective Theories of Strings}
\setcounter{equation}{0}
\noindent
The radius of the circle of the previous example was taken to be a
(fixed) constant. In string theory however it can vary, and it really
corresponds to
a field.
Let us consider a compactification of string theory, where part of
the internal space is a circle with radius $R^2>>\alpha'$.
As mentioned before, the low energy spectrum is composed of momentum
modes
only.
Thus when we derive the low energy effective field theory we must
integrate
out the heavy (winding) modes.

Imagine now following this effective field theory, as the radius
becomes smaller.
It is obvious that when we reach the region with $R^2\sim \alpha'$ we
will encounter some strange behavior, namely non-unitary and/or
singularities.
The reason is that at this region we have integrated out fields that
have
comparable mass to the ones we kept, and this is inconsistent.
There is no singularity in the full string spectrum, just our
approximation
broke down.

Such a change of $R$ can happen in a cosmological context.
There are exact solutions in string theory where $R$ changes with
time
\cite{kk}.
The functional behavior can have the forms
\be
R(t)\sim t\;\;\;,\;\; R(r)\sim 1/t\;\;,\;\;R(t)\sim \tan(at)
\ee
\be
R(t)\sim \coth(at)\;\;,\;\;R(t)\sim \tanh(at)
\ee

Thus we have an example of a cosmological situation where a single
effective field theory is not enough to describe the entire evolution
of the universe.

\section{Topology Change and Black Hole Singularities}
\setcounter{equation}{0}
\noindent
As we mentioned earlier not all string solutions have a conventional
geometrical interpretation.
Here we will examine situations in which the geometry of 3-space
changes as a function of time.
Of particular interest are situations in which at early times
curvatures
are weak in which there is a well defined geometry, at intermediate
times
curvature gets strong and thus the geometrical interpretation is
strong, and at late times the universe flattens again so that
geometry is again well defined
but can be quite different from the original one.
In such cases 3-d topology can change.

In general relativity, topology cannot change without going through a
singularity.
We will see that in string theory this can happen smoothly.
If however we follow the evolution through the effective field theory
we will encounter singularities in the region of strong curvatures.
As argued however in the previous section this only indicates the
breakdown of the effective field theory.
There is no singularity in the context of the full string theory.
Solutions exhibiting smooth topology change in string theory when
some parameters are varied have described in \cite{agm},\cite{gk}.
Exact solutions for time-dependent topology change have been
described in \cite{kk}.
We will describe here such an example.

At time $t=0$ the 3-space has the topology of a line times a disk.
The disk is not flat and its metric is
\be
ds^2=dr^2+\tanh^2r~d\theta^2\label{m1}
\ee
The space evolves at intermediate time to arrive at $t=\infty$ to a
space with the topology of a line times a cylinder with metric
\be
ds^2=dr^2+\coth^2r~d\theta^2\label{m2}
\ee
Although this metric looks singular, the associated string theory is
regular.

This brings us to the other interesting question, which is concerned
with the nature of singularities in string theory.

There are exact solutions in string theory which from the naive
(effective field theory) point of view have point-like singularities
and associated horizons.
In fact such apparent singularities in string theory can be worse in
the sense
that we can have solutions with singularities on higher dimensional
hypersurfaces.
As we argued above, close to the singularity the curvature is large
so the effective field theory breaks down. Can we use tools like
duality to say something about what is the nature of such black holes
in string theory?
We will show first examples examples of euclidean manifolds which
although geometrically singular are absolutely regular in string
theory.

The first example is what is known as the SU(2)/U(1) space.
Its metric and dilaton are
\be
ds^2=d\beta^2 +\tan^2\beta d\alpha^2\;\;,\;\;\Phi=\log(1+\cos2\beta)
\ee
where $\alpha,\beta\in [0,2\pi]$.
The curvature and dilaton are singular at $\beta=\pi$.
However the $\s$-model associate with this space can be solved
exactly
and all amplitudes are regular.

The next example is given by the metric (\ref{m2}) along with the
dilaton
$\Phi=\log(1-\cosh 2r)$.
Here also the curvature and dilaton are singular at $r=0$.
Although we cannot solve this model exactly it can be shown \cite{gk}
that it is
dual to the one with metric (\ref{m1}) and dilaton $\Phi=\log(1+\cosh
2r)$
which is perfectly regular.
This duality is relevant since analytic continuation of these metrics
gives
an exact solution to string theory where the four space has two flat
directions
while the other two have a metric (in Kruskal coordinates) and
dilaton given by
\cite{w},
\be
ds^2={du~dv\over 1-uv}\;\;,\;\;\Phi=\log(1-uv)
\ee
The Penrose diagram for this black hole is similar to the standard
Schwarschild
black hole.
The duality between the Euclidean spaces we described above has some
peculiar consequences for this black hole.
It interchanges in particular the horizon and the singularity,
\cite{g}.
This implies that the physics here is quite different from a field
theoretic
black hole.

Using our previous experience with the effects of duality we can
speculate about the physics of such a black hole.
An example of how this type of symmetry can affect string
propagation, can be given (heuristically) as  follows \cite{k2}.
Consider a string background which is singular
(semiclassically) in a certain region.
In the asymptotic region, (which is obtained by some
spacetime-depended radius becoming very large), one has quantum
numbers for asymptotic states that
correspond roughly to windings and momenta. Momentum states are the
only low energy states in this region.
An experimenter sitting at the asymptotic region, far away from the
black hole would like to probe its nature.
He can do this  using the low lying (momentum) modes available to
him.
Consider a momentum mode travelling towards the high curvature
region.
Its effective mass starts growing as it approaches large curvatures.
At some point it becomes energetically possible for it to decay to
winding
states which, in this region, start having effective masses that are
lower
than momentum modes.
In such backgrounds (unlike flat ones) winding and momentum are not
separately conserved so that such a transition is possible.
The reason for this is that there is a non-trivial dilaton field and
thus, winding and momentum conservation is broken by the screening
operators
which transfer it to discrete states localized at the high curvature
region.

In fact it is a general property of string theory that classical
singularities have always associated with them states localized at
the singularity.
These can be interpreted as internal states of the (would be)
singularity.
A useful picture here is that of the hydrogen atom where the
localized states are the bound states, while the scattering states
above form the continuum.
Thus we could say  that particles
interact
with such localized states loosing momentum (in discrete steps) and
gaining
winding number.

Once such a momentum to winding mode transition happens in the
strongly curved
region, the winding state sees a different geometry, namely the dual
one and thus continues to propagate further into the strong curvature
region
since it feels only the (weak) dual curvature.

\section{On Cosmological Singularities}
\setcounter{equation}{0}
\noindent
We believe today that our universe underwent a big bang and continued
expending thereafter. We certainly do not trust Einstein's equations
beyond a time of the order of the Plank scale. If we naively
extrapolate however we will find a (time-like) singularity at t=0
where the universe squeezes at zero volume.

Using the duality ideas  we can analyze a similar situation in the
context of string cosmology.

Let us consider an expanding universe in string theory.
For simplicity we will consider the spatial slice to be a three torus
with a time dependent volume.
A solution of this kind (to lowest order in $\alpha'$) is given
by \cite{m}
\be
ds^2=-dt^2+\sum_{i=1}^3 a_{i}^2~ t^{2b_{i}}~d\sigma_{i}^2
\ee
\be
B_{\mu\nu}=0\;\;,\;\;\Phi={1\over 2}(\sum_{i=1}^3b_{i}~-1)\log
t\;\;,\;\;
\sum_{i=1}^3 b_{i}^2=1
\ee
where $\s_{i}\in [0,2\pi]$ and we can choose $b_{i}>0$ so the
universe is expanding.
As $t\to 0^{+}$ the universe is collapsing to zero volume, and there
is a curvature singularity there.\footnote{There are solutions exact
to all orders in $\alpha'$ with a similar behavior, but we chose one
with the simplest interpretation.}
We can also make the solution isotropic by choosing $a_{i}=a$ and
$b_{i}=1/\sqrt{3}$.
\be
ds^2_{iso}=-dt^2+a^2~t^{2\over \sqrt{3}}
{}~\sum_{i=1}^3 ~d\s_{i}^2
\ee
\be
\Phi={\sqrt{3}-1\over 2}\log t
\ee
In this example space is a torus with a radius that changes with
time.
At late times the radius is large and geometry is well defined with
the low lying states being the momentum states.
When however the volume becomes of the order of the string length,
$t\sim a^{-\sqrt{3}}$, the geometrical picture breaks down since
winding states have energies comparable to the momentum states.
If we continued naively to $T\to 0$ we would think that the universe
shrinks to zero volume, with a curvature singularity.

The correct approach however is that for $t<a^{-\sqrt{3}}$ to use the
metric seen by the winding modes that now dominate the low energy
spectrum.
\be
d\tilde s^2_{iso}=-dt^2+a^{-2}~t^{-{2\over \sqrt{3}}}
\left(\sum_{i=1}^3 ~d\s_{i}^2\right)
\ee
\be
\Phi=-{\sqrt{3}+1\over 2}\log t
\ee
For these modes the universe is expanding as $t\to 0$.
Thus the correct picture is the following: As we go back in time the
universe shrinks until it becomes of Plank size at which point it
starts re-expanding
again.
This idea is central in pre-big bang type of cosmologies in the
context of string theory \cite{v}.

\section{Conclusions}
\noindent
We have presented some ideas on how physics concerning gravity, the
structure of spacetime (black holes and cosmological singularities
being the focus)
can be quite different in string theory compared to point-particle
field theory.
We noted in particular the role played by stringy symmetries, known
as dualities in establishing this non-field theoretic behavior.

However it should be obvious that we are in the beginning of a long
road towards establishing string theory (or maybe a variant thereof)
as the theory
that describes nature.
The low energy properties of string theory (in the matter sector) are
now a subject of active investigation and we hope to be able soon to
tell whether
some ground state of string theory, looks like the standard model at
low energy and agrees with the experimental data.

The gravitational sector gives another window both to test the
theory, but also
indicates the existence of a host of new phenomena (some of them
described in this talk) which might indicate that Nature is always
richer than we think.

\vskip 1cm

\centerline{\bf Acknowledgements}
We would like to thank the organizers of the conference for providing
an interesting forum to discuss these ideas.
Their hospitality and support is gratefully acknowledged.
C. Kounnas was  supported in part by EEC
contracts SC1$^*$-0394C and SC1$^*$-CT92-0789.


\begin{thebibliography}{9}
\bibitem{gsw} M. Green, J. Schwarz and E. Witten, ``{\em Superstring
Theory}", Cambridge Univ. Press, 1987.

\bibitem{ky} K. Kikkawa, M Yamazaki, {\it Phys. Lett.} {\bf B149}
(1984) 357;\\
N. Sakai, I. Senda, {\it Prog. Theor. Phys.} {\bf 75} (1986) 692.

\bibitem{bu} T. H. Buscher, {\it Phys. Lett.} {\bf B201} (1988) 466.

\bibitem{k} E. Kiritsis, {\em Mod. Phys. Lett.} {\bf A6} (1991) 2871.

\bibitem{rv}  M. Ro\v cek, E. Verlinde, {\em Nucl. Phys.} {\bf B373}
(1992) 630.

\bibitem{gr} A. Giveon, M. Ro\v cek, {\em Nucl. Phys.} {\bf B380}
(1992) 128.

\bibitem{k1} E. Kiritsis, {\em Nucl. Phys.} {\bf B405} (1993) 109.

\bibitem{kk} E. Kiritsis and C. Kounnas, {\em Phys. Lett.} {\bf B331}
(1994) 51.

\bibitem{agm} P. Aspinwall, B. Greene, D. Morrison, {\em Phys. Lett.}
{\bf B303} (1993) 249;\\
E. Witten, {\em Nucl. Phys.} {\bf B403} (1993) 159.

\bibitem{gk} A. Giveon, E. Kiritsis, {\em Nucl. Phys.} {\bf B411}
(1994) 487.

\bibitem{w} E. Witten, {\em Phys. Rev.} {\bf D44} (1991) 314.

\bibitem{g} A. Giveon, {\em Mod. Phys. Lett.} {\bf  A6} (1991) 2843.

\bibitem{k2} E. Kiritsis in the proceedings of the {\em EPS 93
conference}, Marseille 1993, eds. J. Carr and M. Perrottet.

\bibitem{m} M. Mueller, {\em Nucl. Phys.} {\bf B337} (1990) 37.

\bibitem{v} M. Gasperini and G. Veneziano, {\em Astropart. Phys.}
{\bf 1} (1993) 317.
\end{thebibliography}
\end{document}